\pdfoutput=1
\documentclass[12pt]{article}

\usepackage{cite}
\usepackage{gensymb}

    \usepackage[pdftex]{graphicx}

\usepackage{color}

\newcommand{\bea}{\begin{eqnarray}}
\newcommand{\eea}{\end{eqnarray}}

\begin{document}
\title{ \bf Jets in multiparticle production in and beyond geometry of proton-proton  collisions at the LHC}
\author{M.Yu. Azarkin, I.M. Dremin, M. Strikman}
\maketitle

\begin{abstract}
Experimental findings by CMS on properties of jets and underlying events at high multiplicities in proton-proton interactions at 7 TeV
are interpreted as an indication of increasing role of central collisions with small impact parameters. The value of the average impact parameter of the pp collisions  as a function of the soft hadron multiplicity is estimated. 
 We find an indication that  the rates of 
different hard processes  observed by CMS and ALICE  universally depend on the underlying event charged-particle multiplicity until it becomes four times higher than average. 
It is shown that the increase of the overlap area of colliding protons is not sufficient to explain  the rate of jet production in events with charged-particle multiplicity
 that is more than three times higher than average.  New mechanisms are necessary, like interaction of protons in rare configurations of higher than average gluon density.
Such mechanisms are not included in the present Monte Carlo event generators. Further studies are proposed. 

\end{abstract}

\section{Introduction}
\label{Sec:intro}

Multi-particle production in proton-proton (pp) collisions is governed by several mechanisms of hadron dynamics.
The geometry of the collision plays a crucial role in the relative contributions of these mechanisms.
Both non-perturbative and perturbative QCD mechanisms contribute to the hadron production.

At the parton level, each event of the inelastic particle production is treated as a combination of hard and soft parton-parton interactions plus partonic 
remnants of colliding protons. The hard parton-parton interactions result in jets, which appear in final state as well-collimated bunches of hadrons.
At large enough transverse momenta of jets, they can be treated within  perturbative QCD. The softer components, including also soft ingredients of jets, combine in 
the so-called underlying event (UE). The interplay between soft and hard contributions certainly depends on the  collision energy and on the impact parameter of the collision. 
The complicated structure of the interaction region was discussed in many papers and, in particular, in~\cite{MPI1, MPI2}.  

Several important features of inelastic pp interactions emerge from the analysis of the data on elastic pp scattering (using $s$-channel unitarity)
and analysis of the transverse parton spread as extracted from  hard {\it exclusive} processes like $\gamma + p\to J/\psi \, +p$.
First, one finds that  in interactions at the LHC energies protons are completely absorptive at small central area and have a large semi-transparent peripheral zone (Sec.~\ref{Sec:elastic_a}). 
Second, one finds that partons with large fractions of proton energy ($x$ above $10^{-3}$ ) are concentrated  in the central  absorptive area. 
A detailed and up-to-date  review of the two-scale picture  of proton is given in~\cite{MPI2}.

The goal of the paper is to derive information about  dependence of hadron production  on the impact parameter using the observables, 
exploiting the fact that the relative importance of hard and soft interactions strongly depends on the impact parameter. 
We use the most recent experimental studies of processes with a hard trigger and high multiplicities. 
We  determine up to which maximum  charged-particle multiplicities, the impact parameter picture works and where other mechanisms start to dominate.
An important tool in our studies is the ratio of the multiplicity of the hard subprocesses for a given range of multiplicity to the one in bulk of inelastic events.  
We demonstrate that this ratio exhibits  {\it universality pattern} when plotted against $N_{\rm ch}/\left<N_{\rm ch}\right>$. 
This result is practically the same for jet production with $p_{\rm T}>$~5~GeV/{\it c}  and  $p_{\rm T}>$~30~GeV/{\it c}, 
as studied by  CMS  for associated charged particles detected in $\left|\eta \right|< 2.4 $ range. 
Moreover, the universality holds when we compare the CMS ratios  with those reported by ALICE for $J/\psi$, D, B-meson  production~\cite{HardScales} in a factor of 3 smaller $|\eta|$ interval. 
We also argue that a new regime sets in at $N_{\rm ch}/\left<N_{\rm ch}\right> \geq 3$, corresponding to very central collisions 
in which the rate of the hard-probe multiplicity exceeds the maximum value allowed by the geometry of the collision.
This new regime may  correspond to selection of configurations in the colliding nucleons with larger than average relatively-small-$x$ ($x\sim10^{-3}-10^{-2}$) gluon density.

The paper is organized as follows. In Sec.~\ref{Sec:elastic} we review the  transverse geometry of bulk  and hard-probe triggered pp collisions. 
Next, in section~\ref{Sec:Jets}, we analyse some recent LHC results ~\cite{UE_CMS_900, UE_CMS_7000, UE_ATLAS_7000, UE_ALICE_7000} as well 
as more detailed data of CMS collaboration~\cite{FSQ12022} on properties of jets and UE in different charged-particle multiplicity intervals. 
In section ~\ref{Sec:Beyond}, we use jet production as a way to test (calibrate) centrality dependence of events on their multiplicity.
We present our conclusions and suggest strategies for further studies in Sec.~\ref{Sec:Conclusion}.

\section{Geometry of soft and hard pp collisions}
\label{Sec:elastic}

\subsection{Proton structure obtained from elastic scattering}
\label{Sec:elastic_a}

\begin{figure}[h]
\begin{minipage}[h]{0.49\linewidth}
\center{\includegraphics[width=1\linewidth]{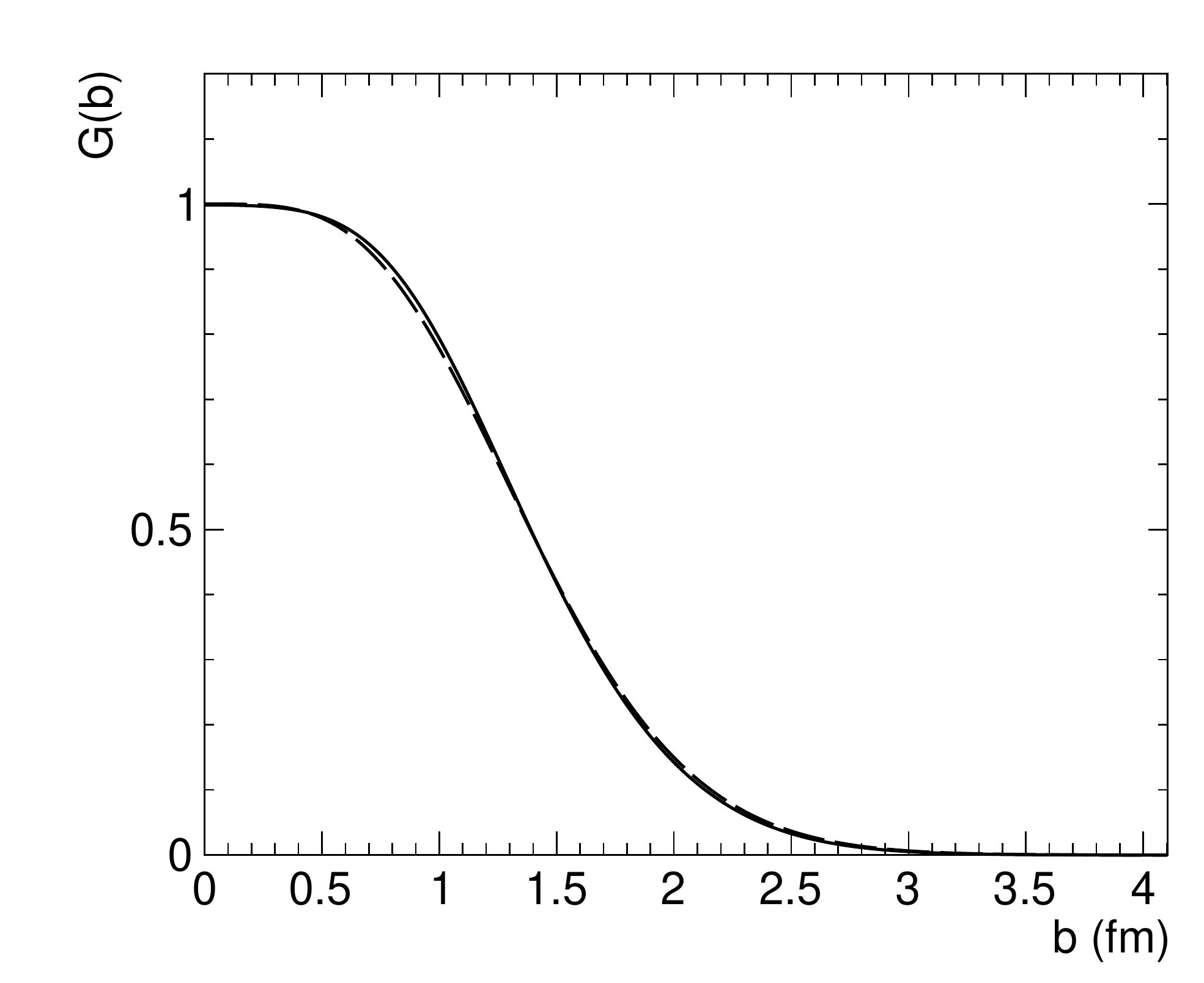} \\ (a)}
\end{minipage}
\begin{minipage}[h]{0.49\linewidth}
\center{\includegraphics[width=1\linewidth]{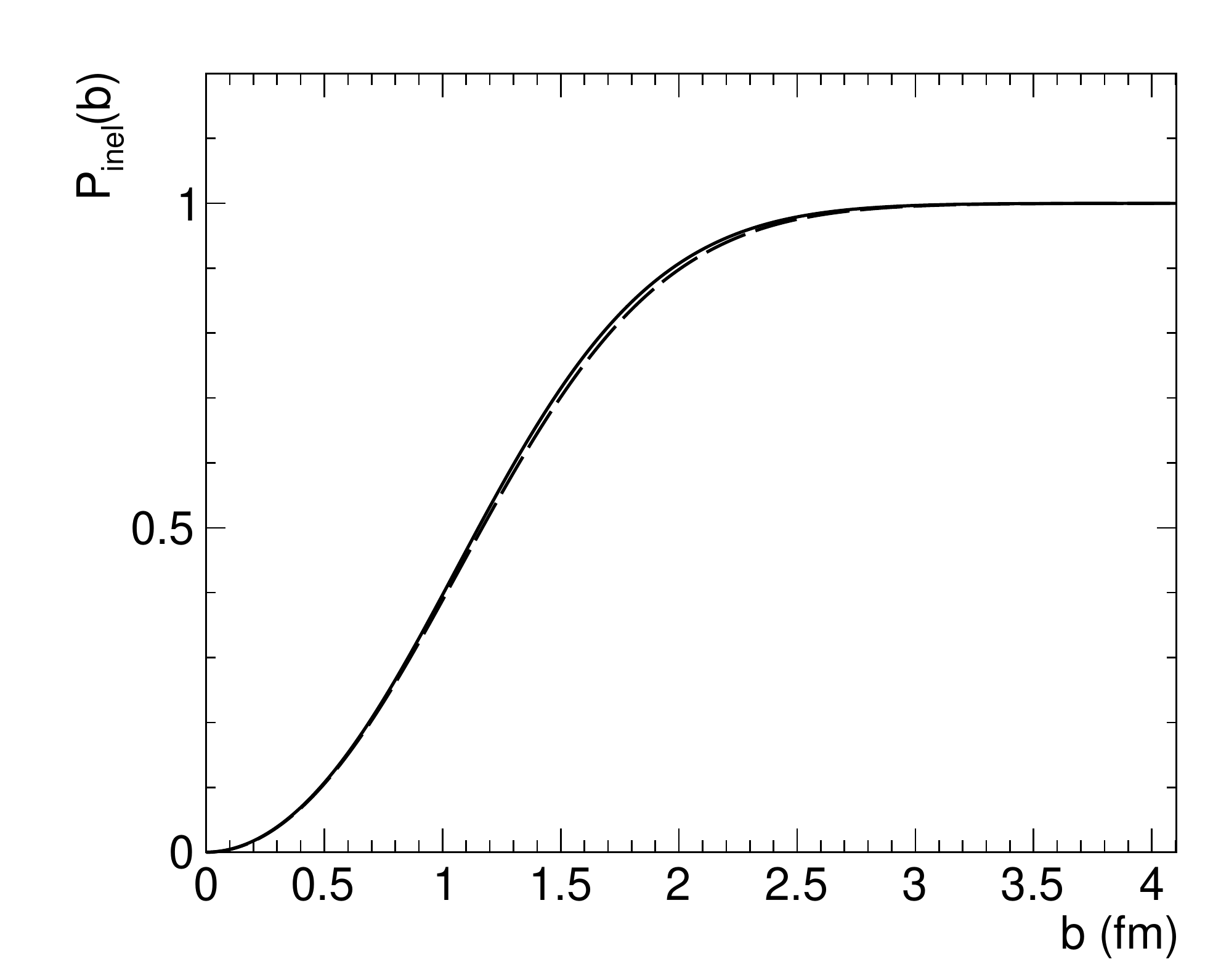} \\ (b)}
\end{minipage}
\caption{Overlap function (a) and probability of inelastic collision with an impact parameter smaller than b (b), according to \cite{dnec}~(solid lines) and \cite{MPI2}~(dashed lines).}
\label{OverlapFunction}
\end{figure}

The impact of the proton structure on inelastic processes can be viewed from
the overlap function defined by the unitarity condition
for elastic scattering amplitudes. It has been directly computed~\cite{dnec}
from experimental data obtained by the TOTEM collaboration
~\cite{totem, diff} for pp scattering at 7~TeV. The corresponding formula in
the impact parameter representation is
\begin{equation}
G(s,b)=2{\rm Re}\Gamma (s,b)-\vert \Gamma (s,b)\vert ^2,
\label{unit}
\end{equation}
where $G(s,b)$ is the overlap function determining the inelastic profile
of colliding protons, and
\begin{equation}
i\Gamma (s,b)=\frac {1}{\sqrt {\pi }}\int _0^{\infty}dqqf(s,t)J_0(qb).
\label{gamm}
\end{equation}
is the Fourier-Bessel transform of the elastic scattering amplitude $f(s,t)$
related to the differential cross section as
\begin{equation}
\frac {d\sigma }{dt}=\vert f(s,t)\vert ^2.
\label{dsigma}
\end{equation}
and normalized as
\begin{equation} 
\sigma _{tot} (s)=\sqrt {16\pi }{\rm Im}f(s,0).
\label{norm}
\end{equation}
The smallness of the real part of $f(s,t)$ corresponding to small
${\rm Im}\Gamma (s,b)$ implies that one can compute $G$ with high precision
either directly from experimental results (see Fig. 3 in~\cite{dnec}) or
assuming the Gaussian profile of the elastic contribution $\Gamma (s,b)$
(see Fig. 4 in~\cite{MPI2}). The obtained shapes of $G(s,b)$  (see Fig.~\ref{OverlapFunction}~(a)) are similar and
show the pattern with rather flat shoulder at small impact parameters $b$ with
subsequent quite steep fall-off. Attempts to fit it by a Gaussian fail because
the plateau up to $b\approx 0.4 - 0.5$~fm is very flat. This is discussed in
more details in \cite{drjl}.

When compared to ISR results~\cite{dnec}, the overlap function and,
consequently, the blackness (opacity) of protons at 7~TeV somewhat increases
in the central region approaching complete saturation. At the same time, a much
stronger increase, about 40~$\%$, is observed in the peripheral
region near 1~fm. Therefore, the periphery starts to play an increasing role in
multiparticle production.

To illustrate the interplay between the black and gray regions we compute the relative contribution of the impact parameters smaller than $b$ to the inelastic cross section:
 \begin{equation}
P_{inel}(b)=\int _0^b~d^2b G(s,b) /\sigma _{\rm inel}(s),
\end{equation}
where $\sigma _{\rm inel}$ - inelastic cross section of pp collisions.
 One can see from Fig.~\ref{OverlapFunction}~(b) that the main 
contribution to the inelastic cross section originates from the gray area while the dark region ($b\le  0.4$ fm) constitutes only about 8~\%.

\subsection{Geometry of dijet production}
\label{Sec:elastic_b}

The transverse distribution of partons in nucleons is given by the generalized parton distributions $f_j(x,Q^2, t)$, which are measured in exclusive hard processes.
In these processes we consider $Q \sim p_{\rm T}^{\rm jet}$. Their Fourier  transform,  $f_j(x,Q^2,\rho)$, determines  the geometry of the inclusive hard interactions\cite{Frankfurt:2003td}.
The probability that the dijet collision occurs at a given $b$ is 
\bea
P_2 (x_1, x_2, b|Q^2) &\equiv&
\int \! d^2\rho_1 \int \! d^2\rho_2 \; 
\delta^{(2)} (\bf{b} - \bf{\rho}_1 + \bf{\rho}_2 )
\nonumber \\
&\times& F_g (x_1, \rho_1 |Q^2 ) \; F_g (x_2, \rho_2 |Q^2) ,
\label{P_2_def}
\eea
where $\rho_{1, 2} \equiv |{\bf \rho}_{1,2}|$ are the transverse distances of the two partons from the center of their parent protons~\cite{Frankfurt:2003td}
and the relation $f_j(x,Q^2,\rho)=f_j(x,Q^2) F_j(x,\rho |Q^2)$ holds.
One finds that the transverse spread of $F_g(x,\rho |Q^2)$ slowly increases with decrease of $x$ 
at fixed $Q^2$ and slowly decreases with increase of $Q^2$ for fixed $x$. 

\begin{figure}[hbtp]
\begin{center}
\includegraphics[ width=0.5\textwidth]{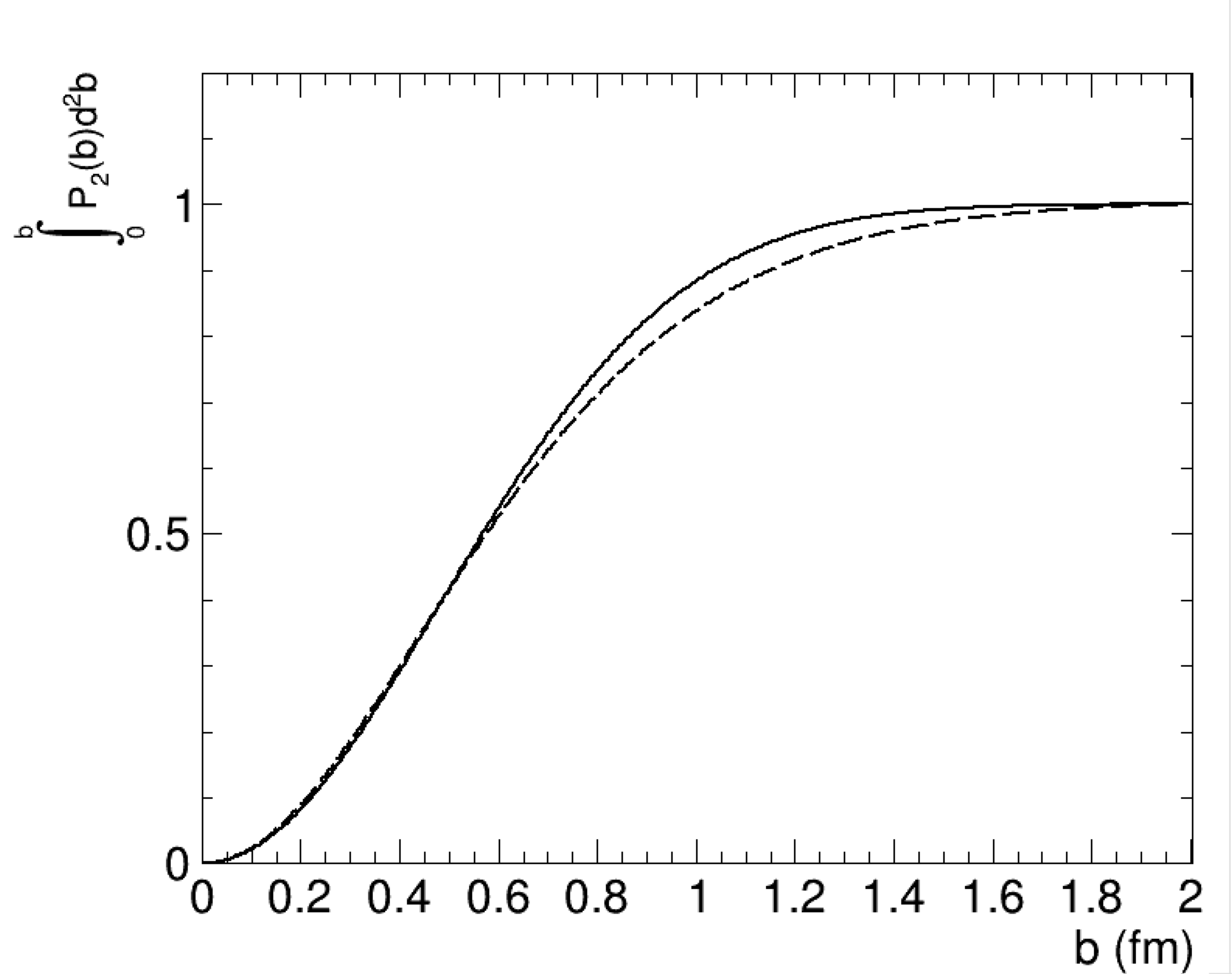}%
\caption{Fraction of the inclusive jet cross section originating from impact parameters from 0 to b.
Solid and dashed lines represent two parameterizations  of $P_{2}(b)$ as given by Eq.~(11) in~\cite{MPI2}.}%
\label{P2_b_int}%
\end{center}
\end{figure}

The distributions of probabilities of hard and soft interactions are compared in Fig.4 in~\cite{MPI2}. One can see that  $b$ distribution for hard processes is much more narrow
than for bulk events. This is reflected in the probability of small $b$ for hard collisions (see plot of $\int_0^b d^2b P_2(b)$ in Fig.~\ref{P2_b_int}) to be much higher
 than for the bulk inelastic collisions (Fig.~\ref{OverlapFunction}b).

\section{Jet and UE}
\label{Sec:Jets}

In this section, we show the connection between standard UE studies and recent study of jet properties as  a function $N_{\rm ch}$, 
exploiting only variables used in these studies, and propose the way to its interpretation. 
In Sec.~\ref{Sec:Beyond}, we will elaborate  on the interpretation and substantiate it with more experimental data. 

The UE properties are usually studied with reference to the direction of the particle or of the jet with largest  $p_{\rm T}$. 
Both approaches have advantages and disadvantages.  The leading jet is more directly related to the initial parton, however it can be affected 
by UE contribution. While the leading charged particle is less related  with the parent parton,  it is not affected by UE. Usually, three distinct topological regions in the hadronic 
final state are thus defined by the azimuthal angle difference $\Delta \phi$ in the plane transverse to the beam between the directions
of the leading object (particle or jet) and other hadrons. Hadron production in the Ònear-sideÓ region with  $|\Delta \phi|<60\degree$ and in the 
Òaway-sideÓ region with $|\Delta \phi|>120\degree$ is expected to be dominated by the hard 
parton-parton scattering and radiation. Thus, UE structure can be best studied in the ÒtransverseÓ region 
with $60\degree<|\Delta \phi|<120\degree$.

\begin{figure}[hbtp]
\begin{center}
\includegraphics[ width=0.5\textwidth]{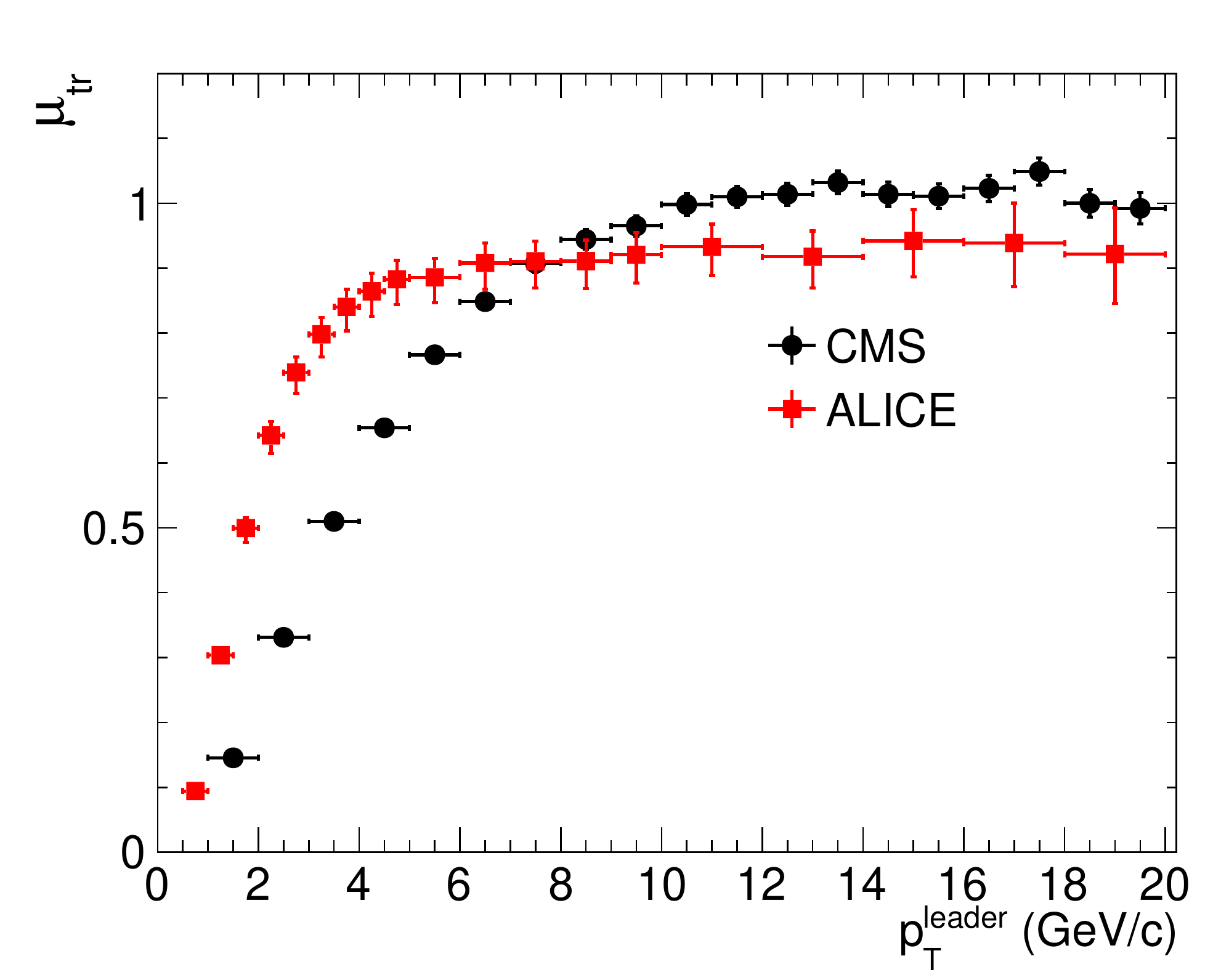}%
\caption{Charged-particle density in the transverse region as a function of $p_{\rm T}$ of leading object (CMS - charged-particle jet, ALICE - charged particle). CMS analyses  particles with $p_{\rm T}>$~0.5~GeV/{\it c} and
$|\eta|<$~2.4, ALICE - $p_{\rm T}>$~0.5~GeV/{\it c} and $|\eta|<$~0.8.}%
\label{M_tr}%
\end{center}
\end{figure}

A number of recent experimental UE studies  use the above described techniques~\cite{UE_CMS_900, UE_CMS_7000, UE_ATLAS_7000, UE_ALICE_7000}.  
ALICE and ATLAS collaborations use a leading charged particle as a reference, while CMS collaboration uses a leading charged-particle jet. Despite very 
different pseudorapidity ranges of ALICE and ATLAS experiments ($\eta<$0.8 and $\eta<$2.5), their results are very close.  Of particular interest is 
particle density in transverse region defined as follows:
\begin{equation}
\mu_{\rm tr} = \frac{N_{\rm ch}^{\rm tr}}{\Delta\eta \Delta(\Delta\phi)},
\label{M_tr_formula}
\end{equation}
where $N_{\rm ch}^{\rm tr}$ is the charged-particle multiplicity in the transverse region, $\Delta\eta$ is the pseudorapidity range studied, $ \Delta(\Delta\phi)$ is the azimuthal width of the transverse region.
The transverse charged-particle density as a function of leading object  at $\sqrt(S)=7$~TeV is shown in Fig.~\ref{M_tr}. The dependence saturates at some $p_{\rm T}=p_{\rm T}^{\rm crit}$, which is $\approx4-5$~GeV/{\it c}  
and $\approx 8$~GeV/{\it c} for leading charged particle and leading charged-particle jet techniques, respectively. According to the two-scale picture of the proton structure 
described in Sec.~\ref{Sec:intro} and elaborated in Ref.~\cite{MPI2}, the observed plateau corresponds directly to the plateau in Fig. 1(a) at $b\le \mbox{1.0 fm}$ because larger $p_T^{leader}$ 
implies smaller $b$ than in minimal bias events and  can be interpreted as an indication of the dominance of the central collisions. 

Since we  aim to reveal a connection between UE studies~\cite{UE_CMS_900, UE_CMS_7000, UE_ATLAS_7000, UE_ALICE_7000}
and studies of jet properties  as function of $N_{\rm ch}$~\cite{FSQ12022}, we need to estimate  the $N_{\rm ch}$ that corresponds  
to the plateau of transverse multiplicity density. The $N_{\rm ch}$ in ~\cite{FSQ12022} is defined as a total number of charged particles with $\eta<2.4$ and $p_{\rm T}>$~0.25~GeV/{\it c}, 
while $\mu_{\rm tr}$  is obtained with charged particles with  $p_{\rm T}>$~0.5~GeV/{\it c}.
From Fig.~\ref{M_tr}  we can see that the transverse multiplicity density  saturates at  $\mu_{\rm tr}^{\rm sat} \approx$~1.0. 
The charged-particle multiplicity of UE can be roughly estimated in assumption of flat $\eta$-distribution as follows: 
\begin{equation}
N_{\rm ch}^{\rm UE} = \mu_{\rm tr} ^{\rm sat}\delta\eta\delta\phi \approx 30,
\label{NchUE}
\end{equation}
where $\delta\eta$ =4.8, $\delta\phi$ = $2\pi$ are the pseudorapidity and azimuthal angle ranges used in ~\cite{FSQ12022}.
Moreover, one should account for  different  $p_{\rm T}$ cuts of charged particles. The correspondence of UE charged-particle multiplicities for different 
$p_{\rm T}$ thresholds is obtained using {\sc  pythia 6 z2*} simulation, which describes UE properties quite well:
\begin{equation}
N_{\rm ch}^{\rm UE}(p_{\rm T}>0.25~{\rm GeV}/{\it c}) = 1.9 \cdot N_{\rm ch}^{\rm UE}(p_{\rm T}>0.5~{\rm GeV}/{\it c}). 
\label{NchUE_025}
\end{equation}
Eq.~(\ref{NchUE_025}) gives  approximately 60 charged particles with $p_{\rm T}>0.25$~{\rm GeV}/{\it c}, belonging to the UE when it reaches a plateau. 
To obtain the total charged-particle multiplicity  of an event,  one should account  a jet contribution. According to tables 4 and 5 of  ~\cite{FSQ12022}, 
a jet contains  5 particles on average, and the jet rate for $50<N_{\rm ch}\le80$ is $\approx 1 $~jet per event. The second, recoiled jet is usually wider and consists of softer particles, 
and thus may be not found by a jet finding algorithm. This is clearly seen in Fig.~2 of Ref.~\cite{UE_CMS_900}. 
Therefore, we conclude that at least 10 charged particles come from jets, and the total $N_{\rm ch}$,  where collisions  become central, equals $\approx 70$.

Results on jet production as a function of $N_{\rm ch}$ presented in ~\cite{FSQ12022} substantiate the analysis of the previous paragraph. 
Indeed,  we see from table 4 of Ref.~\cite{FSQ12022}, that average  $p_{\rm T}$ of jets with threshold of $p_{\rm T}>$ ~5~GeV/{\it c} lies between 7--8~GeV/{\it c}. 
This value matches to the $p_{\rm T}^{\rm crit}$ at which central pp collisions  may occur (see Fig.~\ref{M_tr}).  Therefore,  the jet rate at the thresholds 
5~GeV/{\it c} can serve as a measure of collision centrality. 
From that table, one can see that events starting from $N_{\rm ch}\approx$~60-70 have one jet. 
This means that the most central collision geometry  is reached around these values of $N_{\rm ch}$. 
Other mechanisms should be responsible  for $N_{\rm ch}$ higher than 70, rather than increasing overlapping area of colliding protons. 
Indeed, present MC event generators  completely fail to describe the charged-particle jet rate with $p_{\rm T}$ thresholds of 30~GeV/{\it c} for $N_{\rm ch}>70$ (Fig. 7 of ~\cite{FSQ12022}).
We conclude, that the estimate of $N_{\rm ch}$ of central events obtained from UE  studies is consistent with the value obtained from the jet studies as a function of  $N_{\rm ch}$.

\section{The impact parameter dependence and beyond}
\label{Sec:Beyond}

\begin{figure}[hbtp]
\begin{center}
\includegraphics[ width=0.5\textwidth]{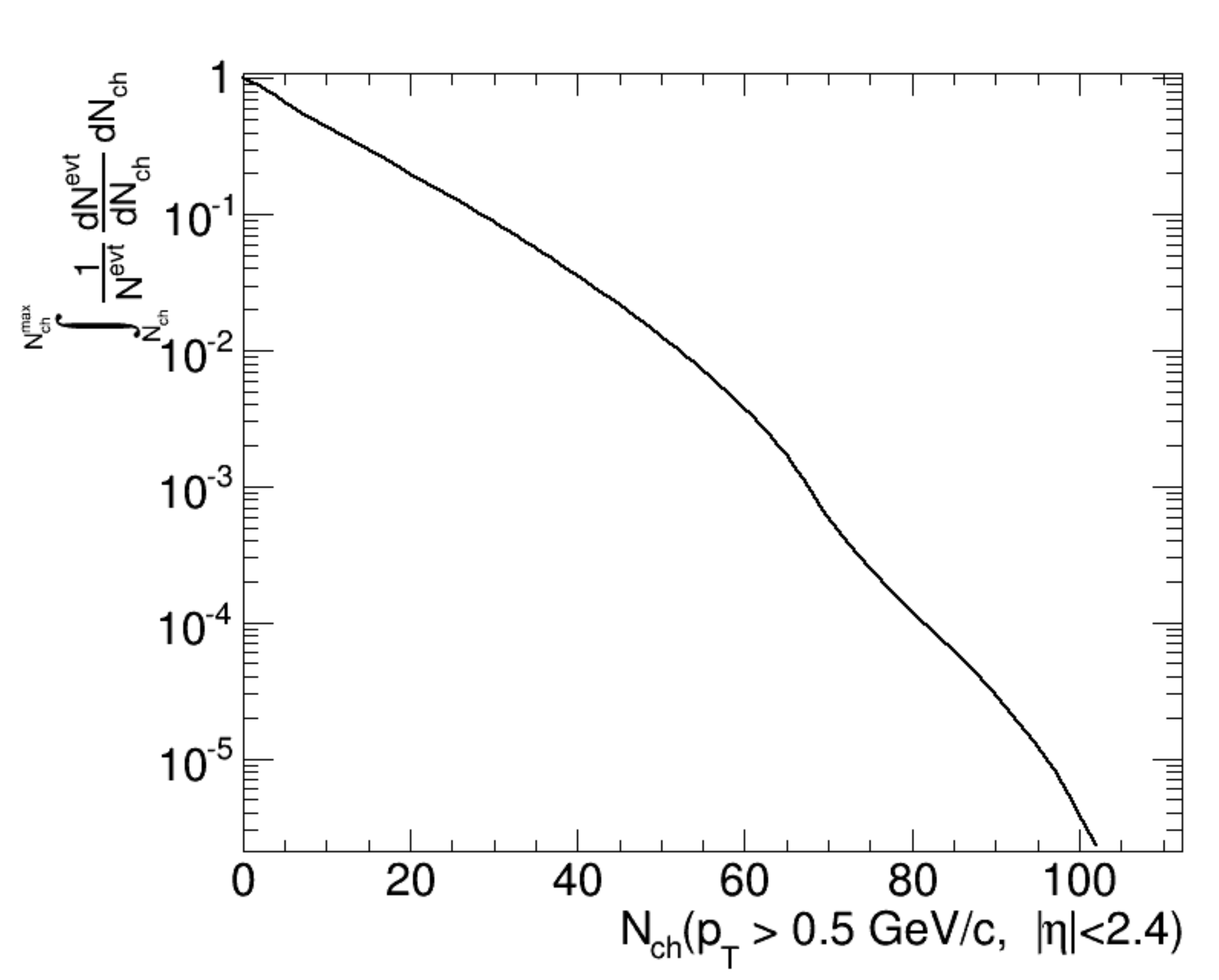}%
\caption{Fraction of events with  $N_{\rm ch}>N_{\rm ch}^{\rm fixed}$. The $N_{\rm ch}$ is defined  as a number of stable charged particles with $p_{\rm T}>$~0.5~GeV/{\it c} and $|\eta|<$~2.4.
The $N_{\rm ch}$ distribution is taken from~\cite{cmsmult}. }%
\label{N_ch_distr_int}%
\end{center}
\end{figure}

\begin{figure}[hbtp]
\begin{center}
\includegraphics[ width=0.5\textwidth]{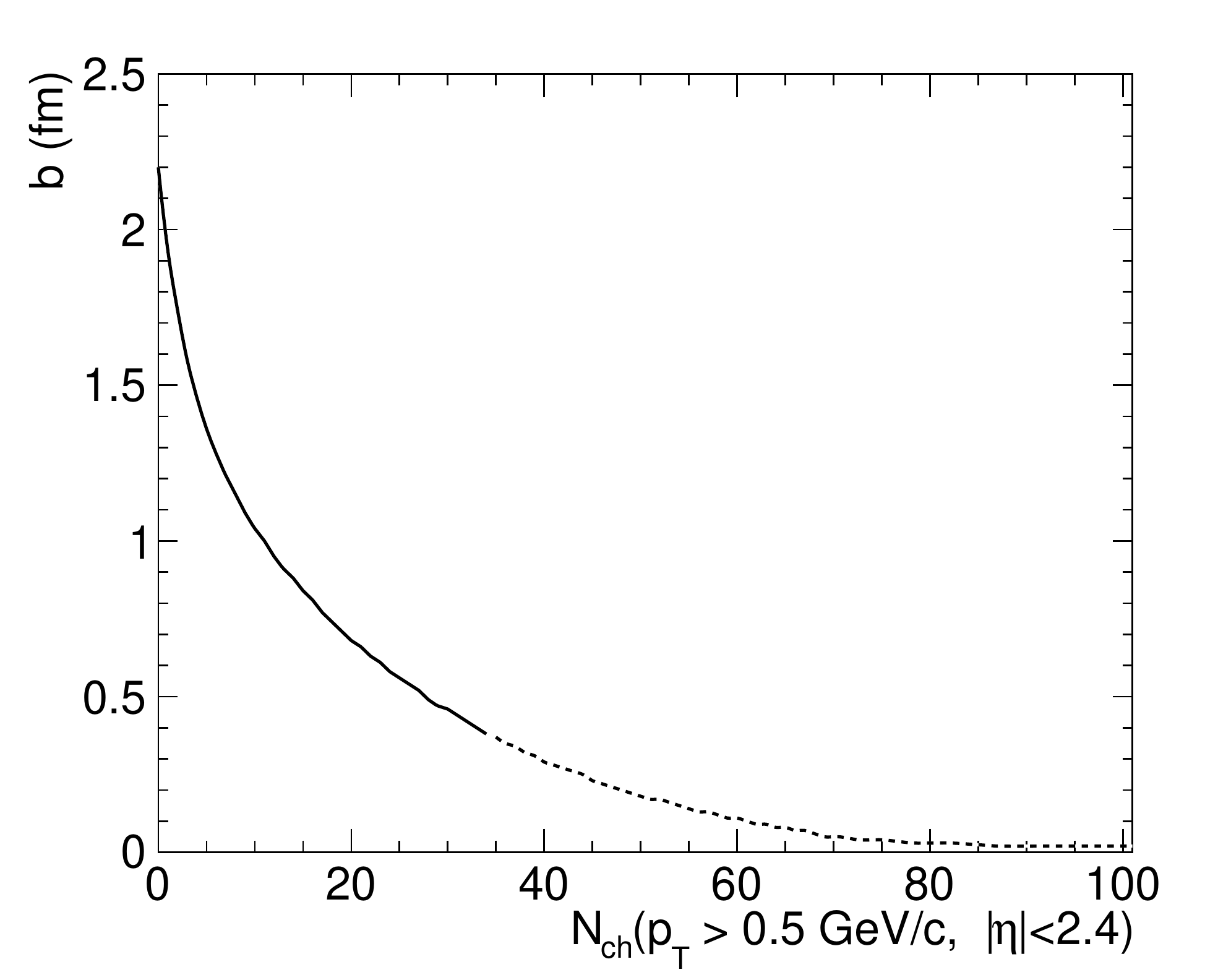}%
\caption{Correspondence between impact parameter and $N_{\rm ch}$. $N_{\rm ch}$ is defined here as a number of charged particles with $|\eta|<$~2.4 and  $p_{\rm T}>$~0.5~GeV/{\it c}.
Since events  with $N_{\rm ch}>$~35 are effectively central as shown below, the correspondence  is not valid there.}%
\label{b_vs_Nch}%
\end{center}
\end{figure}

In Sec.~\ref{Sec:elastic_b}, we have demonstrated that the probability of the central collisions is rather small. It is instructive to compare it with the probability
 to have multiplicity larger than given $N_{\rm ch} $, shown in Fig.~\ref{N_ch_distr_int}. From the comparison of this plot with the plot 
for the probability distribution of inelastic events over $b$ (see Fig.~1(b)), one can make the correspondence between average impact parameter and $N_{\rm ch}$~(Fig.~\ref{b_vs_Nch}).   
We see, that events with $N_{\rm ch}({p_{\rm T}>0.5~{\rm GeV}/c}, |\eta|<2.4)\ge 35 $ mostly originate from collisions with $b \le 0.4$~fm.  
Their probability   is  about 5~$\%$. 
However, measured values of  $N_{\rm ch}$ reach $\approx100$. Such high values (3 times higher than for $b=0.4$~fm) 
can not be  produced even in absolutely  head-on collision,  
if one relies merely on geometric arguments, since possible increase of overlapping area does not  exceed ten-percent level.

\begin{figure}[hbtp]
\begin{center}
\includegraphics[ width=0.50\textwidth]{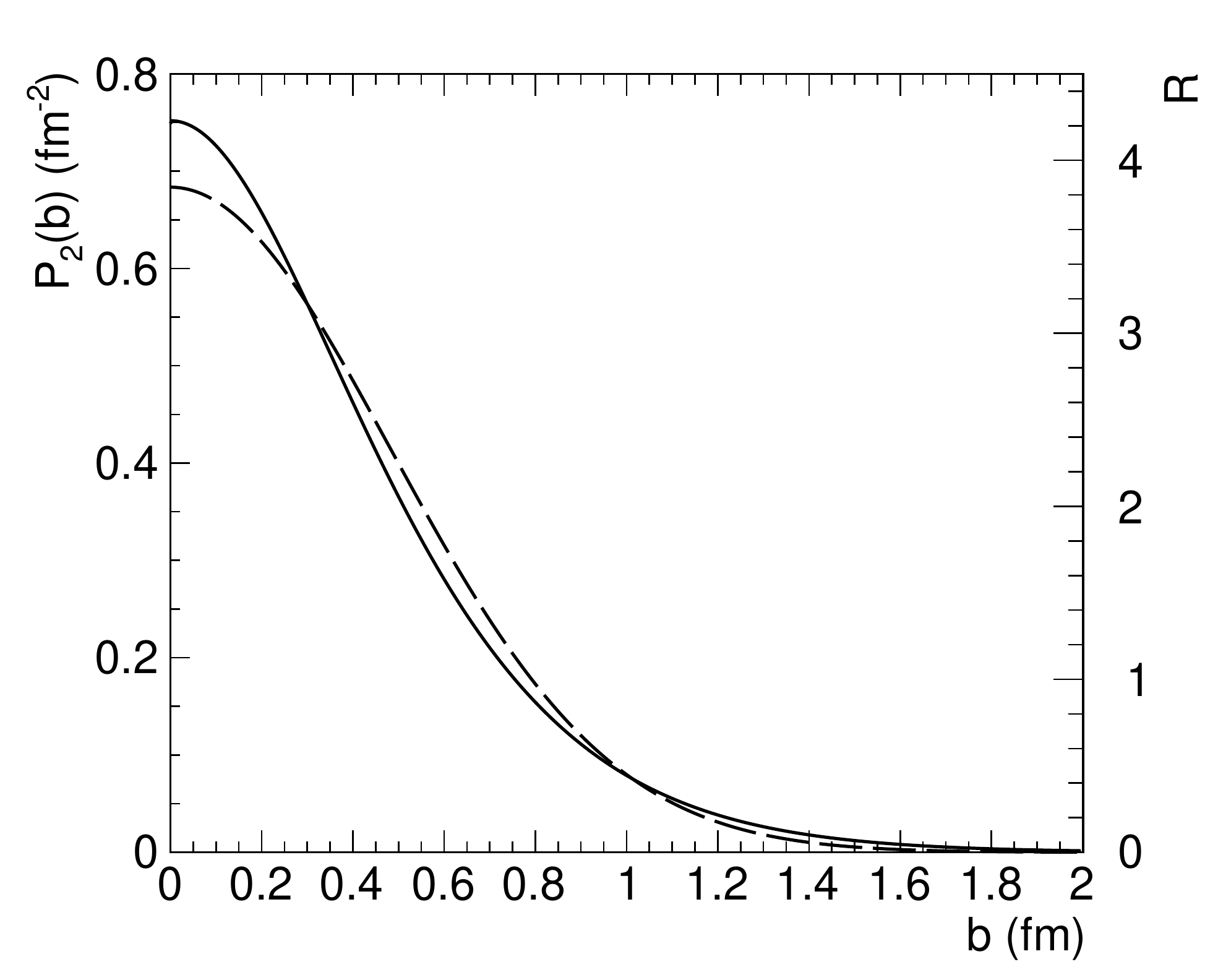}%
\caption{Geometric probability  for two gluons to collide (left y-axis) and inclusive jet production rate with respect to the bulk one (right y-axis). 
Solid and dashed lines represent two parameterizations  of $P_{2}(b)$ as given by Eq.~(11) in~\cite{MPI2}. }%
\label{P2}%
\end{center}
\end{figure}

It was shown in Ref.~ \cite{Strikman:2011ar}, that ratio of the inclusive rate of hard signals at  fixed $b$  to the average one in bulk events is given as follows:
\begin{equation}
R(b)= P_2(b)\sigma_{inel}.
\label{rate}
\end{equation}
We take the inelastic cross-section to include events for which any traces of collision are observed  in the detector (such events are further called minimum bias ).
In particular,  we refer to the CMS experiment, using their phase space. Therefore, inelastic cross-section  taken from~\cite{CMS_inel} $\sigma_{inel}$~=~55~mb 
should be used for further relevant computations. Thus,  a change of the impact parameter of the collision explains values of the ratio (Eq.~\ref{rate}) up to $R\sim 3.8-4.2$ (Fig.~\ref{P2}).
 It is worth noticing that $R$  flattens out already for $b\le 0.3 \div 0.4$~fm.


\begin{figure}[hbtp]
\begin{minipage}[h]{0.49\linewidth}
\center{\includegraphics[width=1\linewidth]{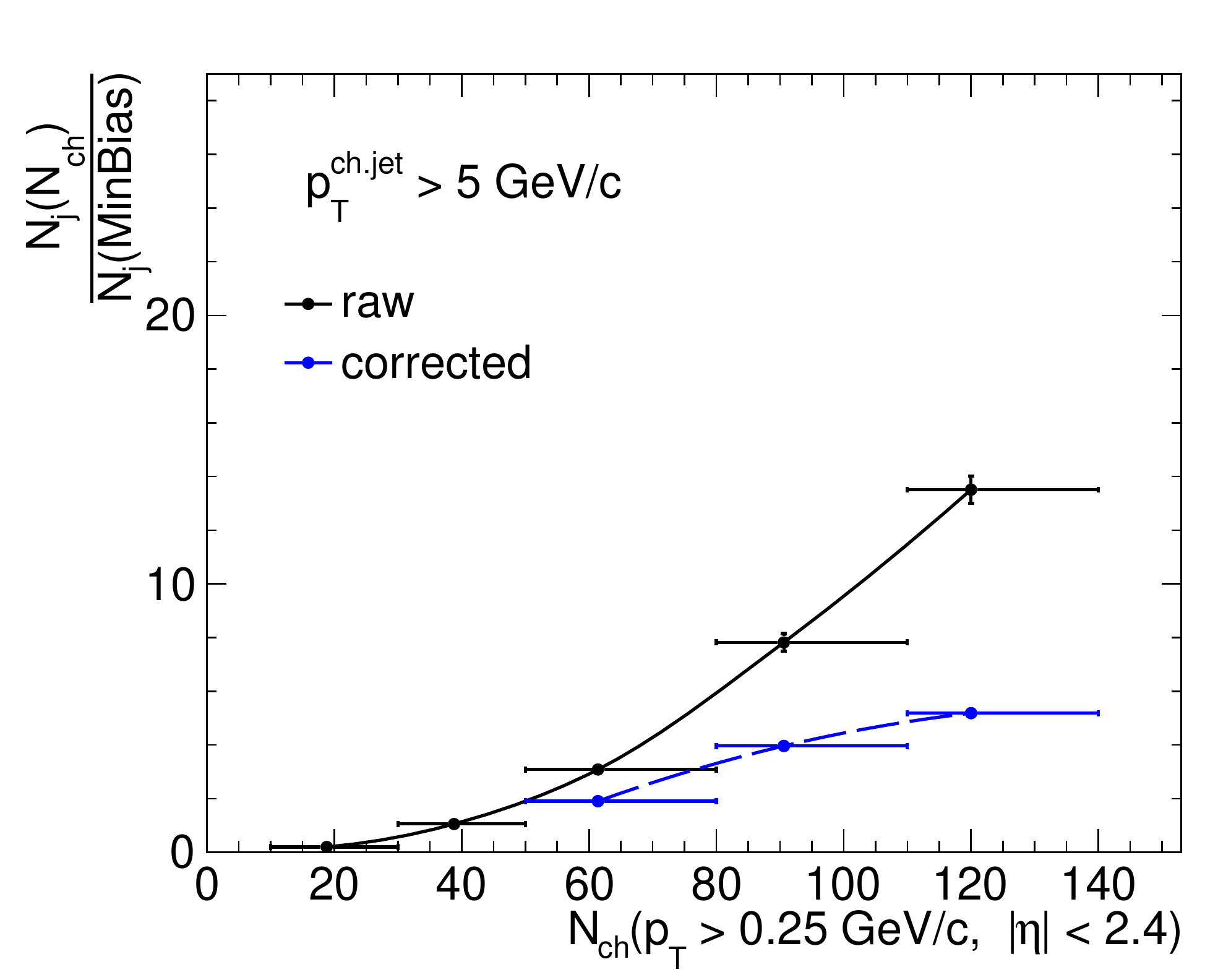} \\ (a)}
\end{minipage}
\begin{minipage}[h]{0.49\linewidth}
\center{\includegraphics[width=1\linewidth]{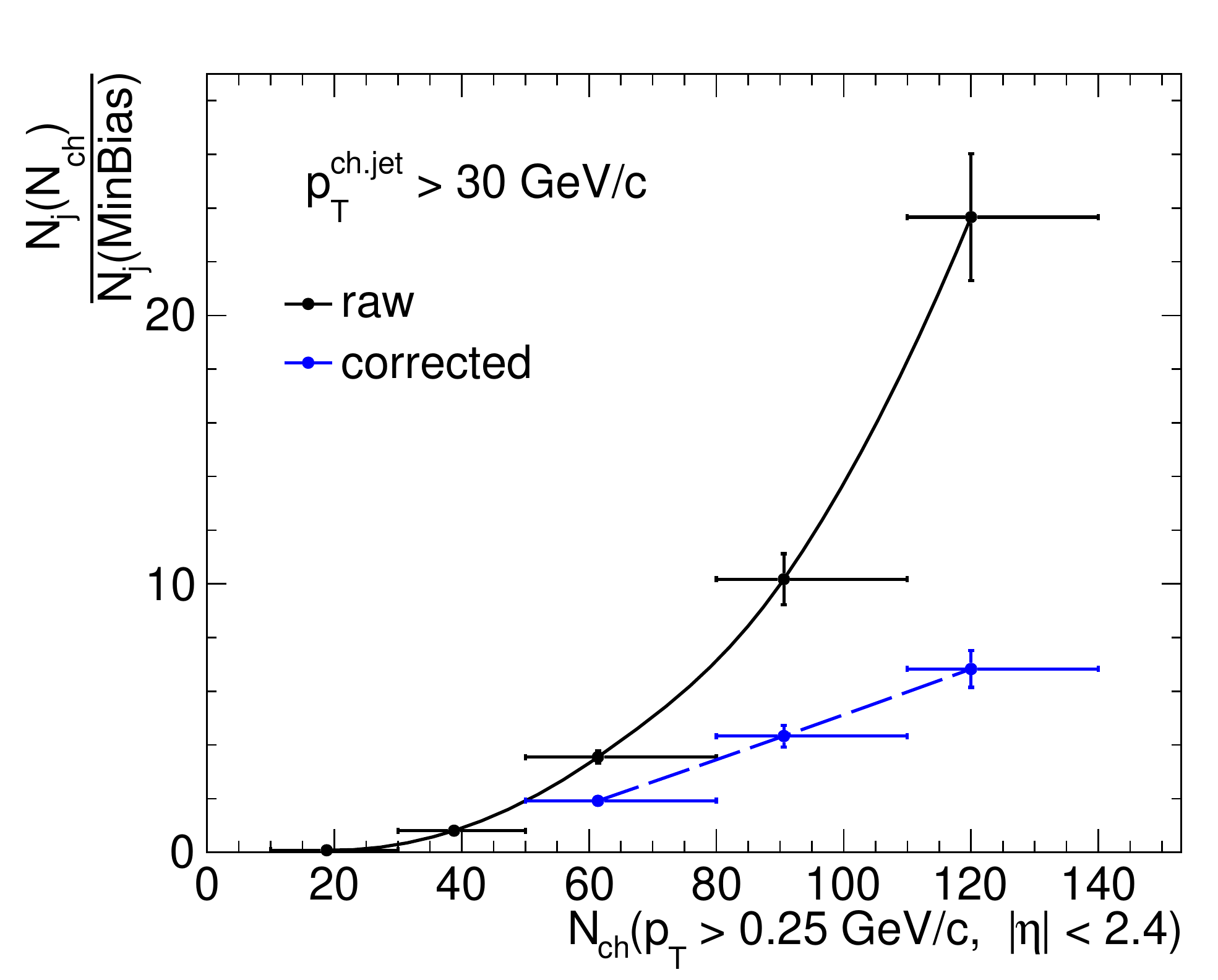} \\ (b)}
\end{minipage}
\caption{Ratio of $N_{\rm j}$  at given $N_{\rm ch}$ to $N_{\rm j} $ of bulk events: (a) - for charged-particle jet $p_{\rm T}^{\rm ch.jet}>$~5~GeV/{\it c}, (a) - for charged-particle jet $p_{\rm T}^{\rm ch.jet}>$~30~GeV/{\it c}. 
The  black solid lines represent  data sorted according to total  $N_{\rm ch}$. Dashed blue lines represent the ratio if the data would be sorted according to $N_{\rm ch}^{\rm UE}$ (note, that the data points are plotted using total  $N_{\rm ch}$). To correct the total  $N_{\rm ch}$ to the $N_{\rm ch}^{\rm UE}$, one needs to subtract $\approx$~10 (15) particles for $p_{\rm T}^{\rm ch.jet}$ threshold of 5~(30)~GeV/{\it c}.}
\label{JetRateRatios}
\end{figure}

We have extracted $R$ for charged-particle jets for two $p_{\rm T}$ thresholds, $p_{\rm T}^{\rm ch.jet} > $~5~GeV/{\it c} and $p_{\rm T}^{\rm ch.jet} > $~30~GeV/{\it c},
from the data taken from Ref.~\cite{FSQ12022} and presented them in Figs. ~\ref{JetRateRatios}(a, b).
The ratios show a very strong increase beyond $N_{\rm ch}\sim 80$.
To compare Eq.~(\ref{rate}) with the data in Figs.~\ref{JetRateRatios}~(a, b)  we need ideally to plot the rate of jet production as function of  $N_{\rm ch}^{\rm UE}$. 
Experimentally,  the purest way to measure the rate is  to select jets produced in one bin of rapidity with multiplicity  measured in another bin of rapidity.
By doing this, we would avoid the contribution of the hadrons produced in the hard trigger component of the event.
In practice, with the current data, we can only try to correct roughly for this effect by using
MC simulations ({\sc pythia~6 ~z2*}) to estimate the average charged-particle multiplicity in the selected jets:  $\Delta N_{\rm ch}$~=~10 (15) for $p_{\rm T}^{\rm ch.jet} > $~5~(30)~GeV/{\it c}. 
Hence, to correct for the jet contribution we need to reduce the experimental ratios by a factor 
$P(N_{\rm ch})/ P(N_{\rm ch}^{\rm UE})$ and plot them as a functions of the $N_{\rm ch}^{\rm UE}$, which is $N_{\rm ch} -\Delta N_{\rm ch}$ here. 
The differential $N_{\rm ch}$ distribution used for computation of the corrections is taken from~\cite{cmsmult}.
Since the low-$N_{\rm ch}$ events have large fluctuations in $\left| \eta \right|$, the correction is not reliable  for $N_{\rm ch}\le$~50 and  the corresponding points are not plotted in Figs. ~\ref{JetRateRatios}(a, b).
One can see that the corrected ratios are approximately the same for two $p_{\rm T}$ cuts. This is consistent with the hypothesis that the rates are determined by the initial state of colliding protons. 

\begin{figure}[hbtp]
\begin{center}
\includegraphics[ width=0.5\textwidth]{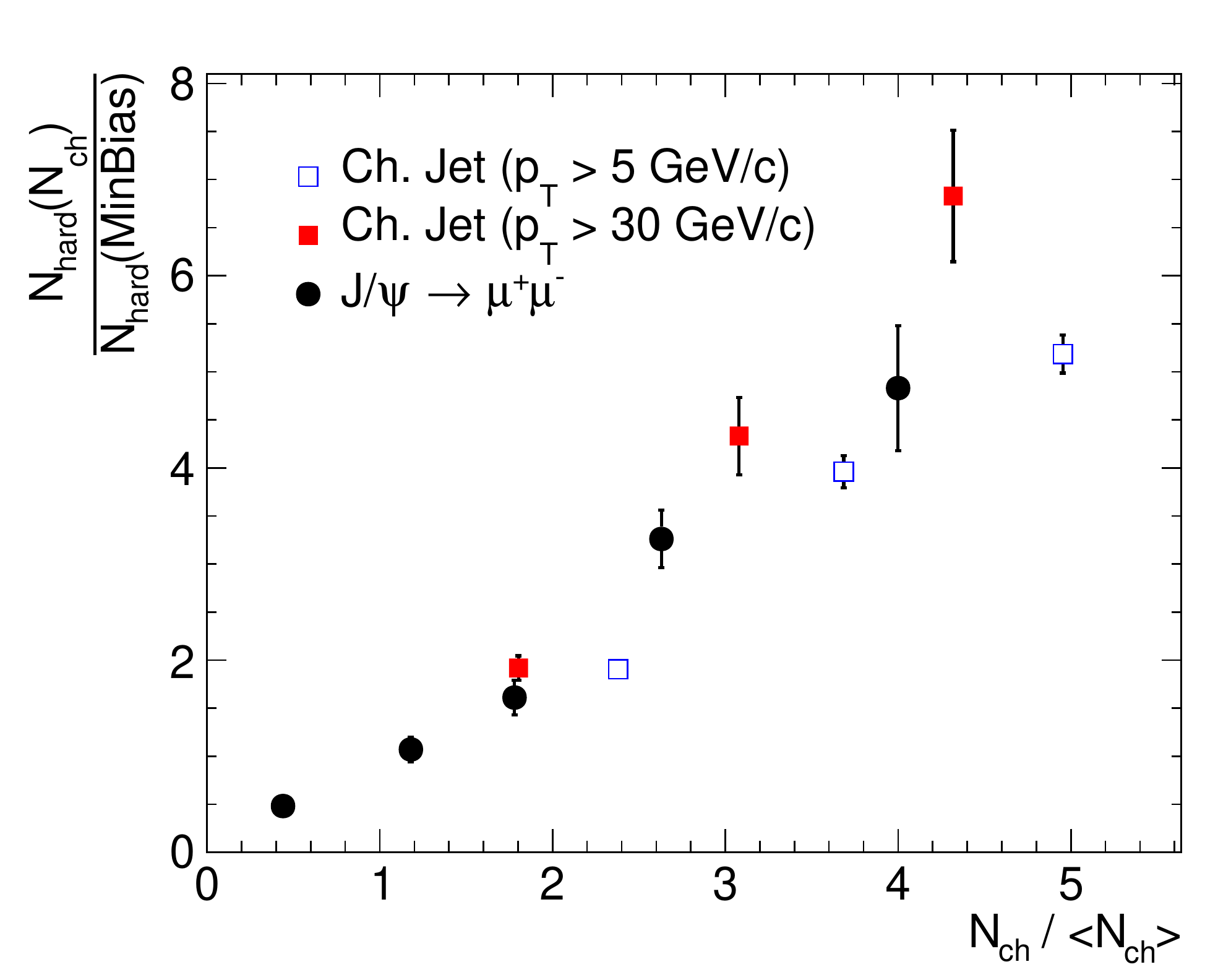}%
\caption{Relative yield  of  hard momentum processes as a function  of  $N_{\rm ch}$, which does not  include particles originating from the hard interactions.}%
\label{HardScaleYield}%
\end{center}
\end{figure}

It is worth noticing, that ALICE has performed studies of a similar quantity, $R$,  i.e. the ratio of the $J/\psi$ multiplicity as a function of $N_{\rm ch}$ normalized to minimum bias $J/\psi$ multiplicity~\cite{HardScales}. They also reported the same ratio for $D$ and $B$-meson production. The observed dependences of $R$ on  $N_{\rm ch}/ \left<N\right>$ are very similar to the one we observe after correcting for the jet contribution (Fig.~\ref{HardScaleYield}). It is worth emphasizing here, that similarity  between $R$ in the two measurements is highly non-trivial as the rapidity intervals used for determination of $N_{\rm ch}$ differ by a factor of $\sim$~3.

\begin{figure}[hbtp]
\begin{minipage}[h]{0.49\linewidth}
\center{\includegraphics[width=1\linewidth]{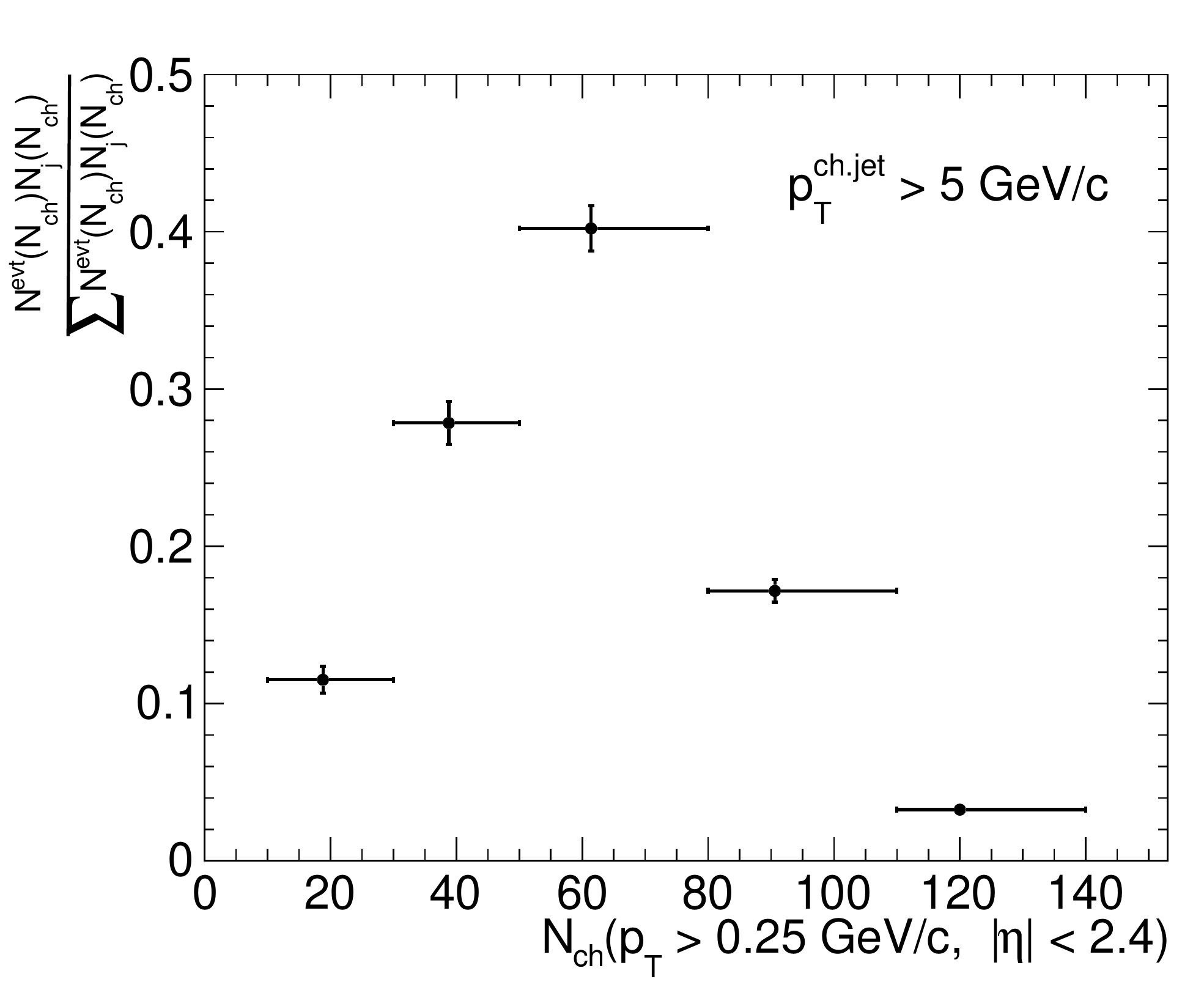} \\ (a)}
\end{minipage}
\begin{minipage}[h]{0.49\linewidth}
\center{\includegraphics[width=1\linewidth]{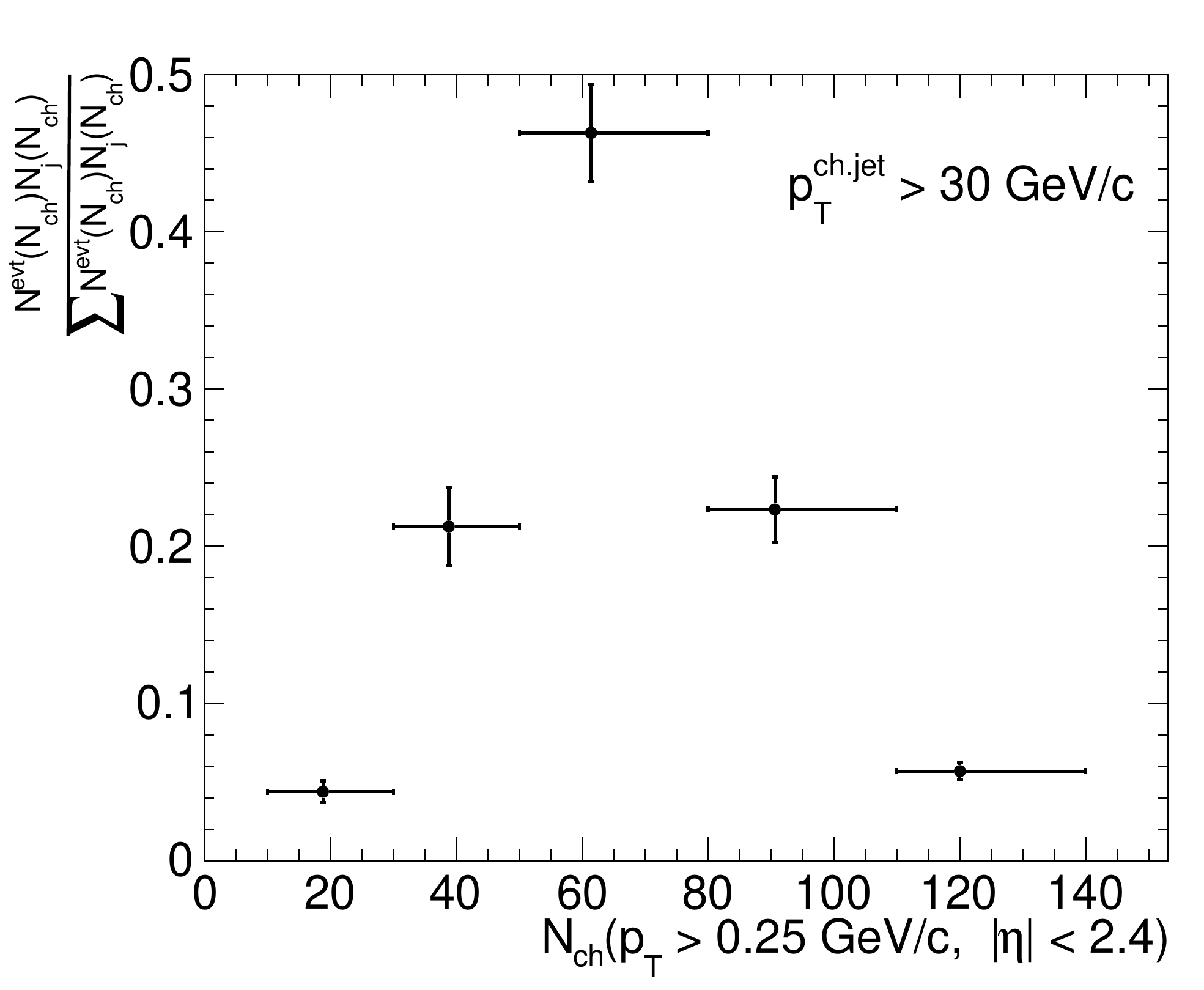} \\ (b)}
\end{minipage}
\caption{Inclusive jet production as a function of $N_{\rm ch}$: (a) - for charged-particle jet $p_{\rm T}^{\rm ch.jet}>$~5~GeV/{\it c}, (a) - for charged-particle jet $p_{\rm T}^{\rm ch.jet}>$~30~GeV/{\it c}.}
\label{InclusiveJetProd}
\end{figure}

We mentioned above that the inclusive rate of the jet production at given $b$ as compared to the bulk rate can be calculated using the information about spatial (transverse)
gluon distributions in nucleons (Eq. (11) of~\cite{MPI2}). 
Hence, it is provocative to consider relative contributions of different bins in $N_{\rm ch}$ to the total inclusive rate of jet production. 
The results are presented in Fig.~\ref{InclusiveJetProd} (a,b) for $p_{\rm T}^{\rm ch.jet} >$~5~GeV/{\it c} (30~GeV/{\it c}).  Since the median of the $P_2$ distribution corresponds to $b \sim $~0.6~fm,  we conclude that  $N_{\rm ch} \sim 60 $ should roughly correspond to that median value of impact parameter. Also, the corrected value of the ratio, $R$,  for third domain, which corresponds to average $b$ of dijet collisions, has  a value of about 2 that is consistent with the expectations of Eq.~(\ref{rate}) for average $b$.

Moreover, the  highest multiplicity points for both $p_{\rm T}^{\rm ch.jet}$ cuts correspond to $R$ well above 4.0 indicating that new mechanisms play a dominant role in this case. 
The rate of jet production in the collision with a trigger  is proportional to the $g_1\cdot g_2/ S$, where  
$g_1$  and $g_2$ are the gluon densities  in the  configurations of nucleons dominating for  a particular trigger and  $S$ is the effective area of overlap, cf. Eq.~\ref{P_2_def}.
Hence, one possibility \cite{Strikman:2011ar} is that the rare high-$N_{\rm ch}$  events are produced in collisions  of protons in configuration with gluon fields, 
that are  significantly stronger than average gluon density\cite{Strikman:2011ar}:
\begin{equation} \frac{ (g_1 g_2/S)_{N_{\rm  ch}/\left< N_{\rm ch}\right> \sim 4 \div 5}}{ \left<g_1 g_2/S\right> }\sim 2
\end{equation}
Probability of such large fluctuations of the gluon field in one or both nucleons is very small -- on the scale of $10^{-2}$, see discussion in  \cite{Strikman:2011ar}. Hence, this scenario  explains the  much smaller probability of these events than the one given by the geometry of the collisions as well as the  similar value of the  enhancement for events with different jet transverse $p_{\rm T}^{\rm ch.jet}$. Another source could be contribution of the higher-order QCD processes, that are not properly accounted or not accounted at all in event generator models. 
However, it is not likely that they would generate the same enhancement for very different $p_{\rm T}^{\rm ch.jet}$ cuts.

Note also,  that  the measured charged-particle multiplicity reaches values of 170-180~\cite{cmsmult}. Then, determining the rate of jet production for these collisions may open a window on the properties of very rare configurations in nucleons.

\section{Summary and conclusions}
\label{Sec:Conclusion}

The role of proton geometry in multiparticle production  at LHC energy has been studied using various experimental data.  
With the help of hard probes we obtain information about the inner regions of the protons. 
At LHC energies, the core of protons became absolutely absorptive due to high parton density. 
Collisions, involving the cores, usually result in high-multiplicity events. Such events contain a large number of high-$p_{\rm T}$ jets.
We have used the dependence of the jet multiplicity on the charged-particle multiplicity  studied in~\cite{FSQ12022} as a tool to look inside proton interaction region. 
We compute  the ratio, $R$,  of the jet multiplicity in the charged-particle multiplicity intervals (corrected for the hard contribution)  to one in minimum bias  events.
The value of the ratio is determined by the geometry of the pp collision up to $N_{\rm ch}/<N_{\rm ch}>  \sim 3.0$, corresponding to the average impact parameters of 0.4 fm.
We have argued that at higher multiplicities one enters a regime of increase of $R$, which is not described by geometry of pp collisions. 
Our analysis of the LHC data strongly suggests that for $N_{\rm ch} > 70$ ( at 7~TeV),   for which the process is dominated by rare central collisions, where colliding protons fluctuate into
special high-density  gluon configurations. This suggests that the events in which ridge was observed prominently also originate from similar collisions.
Also, we find an indication that  the rates of different hard processes observed by CMS and ALICE  universally depend on $N_{\rm ch}$ until it becomes three times higher than average, where geometry effects dominate. This is consistent with the hypothesis that these rates are predominantly determined by the initial state of colliding protons. Moreover, we observe that it holds  even for higher values of $N_{\rm ch}/ \left<N_{\rm ch}\right>$ ( up to 4) for $x$ in the $10^{-2} \div 10^{-3}$ interval. Hence, it would be highly desirable to extend these observations for a wider $x$-range and also 
to study jet production for larger $N_{\rm ch}$ to see whether the jet rate continues to grow, and whether this growth is different for small and moderate $x$ ($x > 0.05$).  Fluctuations of the gluon density may increase relative importance of the multiparton interaction (MPI) mechanism of jet production. 
Thus it would be worthwhile to try to determine the rate of MPI in the high multiplicity events.

More detailed studies are also highly desirable. In particular, it would be preferable to study dependence on the multiplicity of the underlying event rather than on the total multiplicity.
It would be also interesting to study events where multiplicity is significantly smaller than average.  One may expect that these events originate from large $b$ collisions, where the interaction is dominated by the exchange of a single Pomeron. Since properties of the Pomeron do not depend on $b$,  in this scenario $R$ should be practically independent on $\left<N_{\rm ch}\right>$.

Another possible direction  for  experimental studies  would be to investigate how  the  production of leading baryons (for example, production of leading neutrons with $x_F \ge 0.3$) is correlated with $N_{\rm ch}/\left<N_{\rm ch}\right> $. Indeed, it was argued \cite{Drescher:2008zz}   that the  neutron yield in the proton fragmentation region drops with an increase of the centrality of the $pp$ collisions and also decreases  faster  with increase of $x_F$. If so, one  expects that, 
with increase of  $N_{\rm ch}/\left<N_{\rm ch}\right> $, the neutron multiplicity for large $x_F \ge 0.3$  will diminish and it will be 
very strongly suppressed for  $N_{\rm ch}/\left<N_{\rm ch}\right> \ge  4 $. 

\section{Acknowledgments}
M. Azarkin and I. Dremin are grateful for support by the RFBR grants 12-02-91504-CERN-a, 14-02-00099, and the RAS-CERN program.  M. Strikman's research  was supported  by DOE grant No. DE-FG02-93ER40771. we thank T.Rogers for valuable comments. M. Strikman thanks Leonid Frankfurt for numerous discussions and also would like to thank CERN for hospitality, where this work started.


\bibliography{mybib}{}
\bibliographystyle{lucas_unsrt_epjc}

\end{document}